\documentclass[10pt,conference]{IEEEtran}
\IEEEoverridecommandlockouts
\usepackage{graphicx} % Required for inserting images
\usepackage{xspace}
\usepackage{enumitem}
\usepackage{xcolor}
\usepackage{tcolorbox}
\usepackage{framed}
\usepackage[hyphens]{url}
\usepackage[nospace]{cite}
\usepackage{marvosym}

\newcommand{\tool}{\textsc{SCOPE}\xspace}
\newcommand{\one}{({\em i}\/) }
\newcommand{\two}{({\em ii}\/) }
\newcommand{\three}{({\em iii}\/) }

\definecolor{customblue}{HTML}{006ca6}
\definecolor{customgreen}{HTML}{009264}
\definecolor{custombrown}{HTML}{b80d57}
\AtEndPreamble{
\usepackage{hyperref}
 \hypersetup{
  colorlinks = true,
  linkcolor = customblue,
  anchorcolor = purple,
  citecolor = customgreen,
  filecolor = purple,
  urlcolor = black
 }
}

\begin{document}
\title{\tool: Evaluating and Enhancing Permission Explanation Transparency in Mobile Apps}

\author{
\IEEEauthorblockN{Liu Wang$^{1}$, Tianshu Zhou\IEEEauthorrefmark{1}$^{2}$\thanks{\IEEEauthorrefmark{1} Tianshu Zhou is the co-first author.}, Haoyu Wang$^{1}$, Xiyuan Liu$^{3}$, Yi Wang\IEEEauthorrefmark{2}$^{2}$\thanks{\IEEEauthorrefmark{2} Yi Wang (yiwang@bupt.edu.cn) is the corresponding author.}
}
\IEEEauthorblockA{
$^{1}$ Huazhong University of Science and Technology, China
$^{2}$ Beijing University of Posts and Telecommunications, China\\
$^{3}$ Freshippo, Alibaba Group, China\\
}
}

\maketitle

\begin{abstract}
Permission explanations, explanatory text accompanying mobile app permission requests, are crucial for user privacy transparency and informed consent. Despite their importance, current practices often fall short of regulatory expectations due to the lack of systematic evaluation mechanisms. Through an empirical study of 600 mainstream mobile apps, we reveal widespread deficiencies: 15\% of permission requests provide no explanation, others use vague language or technical jargon, and critically, many fail to disclose third-party SDK data access despite these components actively using granted permissions. To address these transparency gaps, we present \tool, an automated multi-agent framework that systematically evaluates permission explanation compliance and generates targeted optimization recommendations. \tool employs four specialized agents working collaboratively: multimodal LLM-based explanation extraction, few-shot learning-based linguistic analysis, dynamic API-based purpose inference, and adaptive report generation. Comprehensive evaluation demonstrates \tool's effectiveness, achieving 98\% accuracy in explanation extraction, 93.5\% consistency in compliance analysis, and 92\% accuracy in purpose inference. A user study with 30 participants shows 84.6\% preference for \tool-optimized explanations, confirming practical utility. Our work provides the first systematic analysis of permission explanation practices and establishes a scalable solution for enhancing mobile app privacy transparency.
\end{abstract}

\section{Introduction}
Mobile apps often require access to substantial amounts of sensitive user information, such as location data, contact lists, and device identifiers, to deliver various functionalities~\cite{cisareport,termsfeed}.
App permissions serve as a critical access-control mechanism regulating an app’s ability to access such sensitive resources: an app needs to request relevant permissions and users need to explicitly grant them before the app can access restricted data or perform restricted actions~\cite{androiddev2}. 
User decisions regarding permission grants–whether to allow or deny these requests–crucially shape the privacy landscape of their devices and associated risks. This is particularly concerning given that a significant proportion of Android apps exhibit permission-greedy behavior, requesting permissions beyond what is necessary for their stated functionality~\cite{betanews,cybernews}.
However, users frequently struggle to make such decisions, mainly because existing practices often fall short in providing adequate transparency regarding permission usage, e.g., how, why, and by whom their data is accessed, leaving users uncertain about whether to give access~\cite{bongard2022necessary}.
Consequently, users often grant permissions simply to ``click away'' permission requests and continue with their primary task, thereby increasing the risk of data leakage and privacy violations~\cite{190982}.

The situation is further complicated by Android’s permission inheritance model, where embedded third-party libraries can access all permissions granted to the host app. 
In modern app development, it is common practice to integrate multiple third-party libraries (e.g., advertising, social, or analytics SDKs) to delegate specific tasks and functionalities, thereby simplifying the development process.
Since these libraries inherit all privileges of the app, they can often become overprivileged relative to their actual functional requirements.
It has been reported the issue of ``permission piggybacking''~\cite{appicaptorblog}, where 
third-party libraries probe already granted permissions and use or collect available data without explicitly requesting permissions themselves.
Moreover, app developers often lack detailed knowledge of the exact internals of each library and its background operations~\cite{Heid_Sonntag_Heider_2025}. 
Consequently, these libraries may collect personal and sensitive user data beyond what is necessary for their primary functionality, creating substantial privacy risks without user awareness or developer intent.

\begin{figure}
    \centering
    \includegraphics[width=0.48\textwidth]{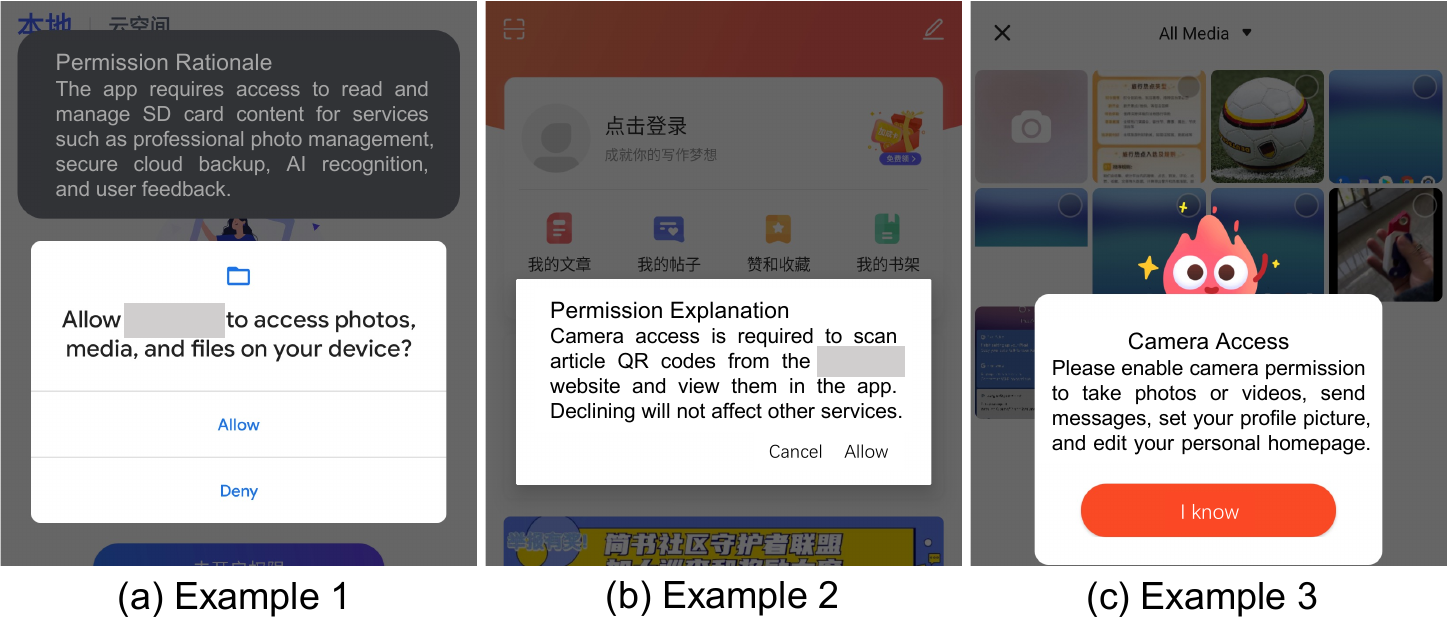}
    \vspace{-0.15in}
    \caption{Examples of permission explanation screens with diverse styles.}
    \vspace{-0.15in}
    \label{fig:examples}
\end{figure}

Privacy regulations such as the EU's General Data Protection Regulation (GDPR)~\cite{GDPR}, the US's California Consumer Privacy Act (CCPA)~\cite{CCPA}, and China's Personal Information Protection Law (PIPL)~\cite{PIPL} emphasize user rights to informed consent and control over personal data, mandating clear, accessible disclosures and explicit consent before data collection. 
While apps commonly rely on privacy policies to communicate data practices, such documents are often lengthy, complex, and difficult for ordinary users to comprehend, thereby diminishing their effectiveness in delivering timely and meaningful transparency. 
Further, the industry has increasingly promoted the use of ``\textbf{permission explanation}'', where developers proactively provide concise and user-friendly in-app prompts that clarify the purpose and necessity of a permission request (as shown in \autoref{fig:examples}) as a mechanism for enhancing user awareness and empowering more informed decision-making.
Given that most users rarely read privacy policies thoroughly and instead rely more on real-time permission prompts for decision-making~\cite{pewresearch,syrenis}, well-designed permission explanations represent a promising pathway toward improved transparency.
Research indicates that users are more likely to provide informed consent when permission requests are accompanied by clear explanations~\cite{274679,nngroup}. 

However, the implementation of permission explanations currently operates with minimal regulatory oversight. Developers have complete discretion over explanation content and design, and no system-level enforcement mechanisms exist to ensure compliance with privacy regulations. 
This regulatory gap creates opportunities for manipulative practices designed to nudge user consent, such as the dark patterns exposed in prior research~\cite{mohamed2024attention,hoober2019mobile}. These concerns underscore the need for systematic evaluation mechanisms to ensure that permission explanations genuinely uphold the transparency principles and user empowerment they are intended to advance.

To understand the current landscape, we first conducted an empirical study examining permission explanation practices across 600 mainstream mobile apps. 
Our investigation revealed three major categories of deficiencies: \one many apps provide no explanation whatsoever, \two others employ vague or technical language that obscures actual data usage, and \three most critically, explanations systematically fail to disclose third-party SDK access despite these embedded libraries inheriting and potentially exploiting all granted permissions. To address these transparency gaps and support developers in creating compliant, user-centered explanations, we present \tool, a multi-agent framework that systematically evaluates permission explanation compliance and generates targeted optimization recommendations. Our framework employs four specialized agents for explanation extraction, linguistic appropriateness analysis, purpose inference, and adaptive report generation, enabling comprehensive assessment across visibility, linguistic quality, and substantive accuracy dimensions.
We conducted extensive evaluations demonstrating that \tool achieves high accuracy in explanation extraction (98\% with multimodal LLMs), linguistic compliance analysis (93.5\% evaluation consistency), and purpose inference (92\% accuracy with contextual enhancement). A user study with 30 participants showed that 84.6\% preferred \tool-optimized explanations over originals, validating the practical utility of our optimization approach. The tool is publicly available~\cite{repo}. In summary, this work makes the following contributions: 

\begin{itemize}
    \item We present the first systematic empirical analysis of permission explanation practices, uncovering widespread transparency deficiencies across three tiers, including prevalent missing explanations, linguistic opacity, and systematic failure to disclose third-party SDK data access.
    \item We introduce a novel multi-agent framework for automated explanation evaluation and optimization. Our approach integrates multimodal LLMs, dynamic API analysis, and regulatory compliance assessment to provide comprehensive, actionable feedback for improving explanation quality across multiple dimensions.
    \item We conduct extensive experimental validation demonstrating both technical effectiveness and user acceptance of our approach. Our evaluation shows superior performance over traditional methods and confirms significant user preference for framework-optimized explanations, validating real-world applicability.
\end{itemize}

\section{Background and Motivation}
\subsection{Permission Explanation}

Since Android 6.0 (Marshmallow) introduced the dynamic permission authorization model, apps must request permissions at runtime for interactive user authorizations. 
Under this model, permission request dialogs are generated uniformly by the Android operating system, typically displaying standardized prompts like “Allow this app to access your location?”. 
Developers cannot customize the content of these system-generated dialogs.
To promote user understanding of permission usage, 
Android officially recommends that developers proactively explain the reasons and necessity for permission requests through in-app interfaces, such as custom dialogs or overlay prompts, before calling the system dialog~\cite{androiddev}. 
Such explanatory information or textual prompt provided by app developers to clarify the purpose and necessity of a requested sensitive permission is referred to as \textit{permission explanation}.

Since developers have full control over the content and design of these explanations, they can vary significantly in format and style.
As illustrated in Figure~\ref{fig:examples}, several design patterns are commonly used: some apps supplement the default system dialog with an additional explanatory text box (Example 1),
others introduce custom explanatory dialogs prior to the system prompt (Example 2), and certain apps adopt more visually engaging designs incorporating illustrations or interactive elements to encourage user comprehension (Example 3). These diverse design practices reflect developers' attempts to balance information transparency with user experience, while also revealing the lack of unified design standards and evaluation criteria in current permission explanation practices.

\subsection{Bright-side and Dark-side}
Permission explanations are widely regarded as a useful mechanism for enhancing user comprehension and informed decision-making, compared to lengthy and often ambiguous privacy policies~\cite{pewresearch,syrenis}. 
Research demonstrated that users were 12\% more likely to grant permissions when they were given a reason for the request~\cite{nngroup}.

However, the actual effectiveness of permission explanations depends heavily on implementation quality, given that developers have full freedom over the content and design in practice. 
While regulatory guidelines provide high-level principles (e.g., data minimization, transparency, and contextual permission requests), they remain conceptually broad and lack specific implementation instructions or concrete templates for developers to follow.
Moreover, there is no system-level enforcement mechanism that prevents developers from violating these regulations.
These lead to inconsistent quality and potential misuse of permission explanations.
For example, Mohamed et al.~\cite{mohamed2024attention} found that real-world implementations of iOS App Tracking Transparency (ATT) alerts employed misleading, vague, or incentivizing language that confused users and distorted their permission understanding.
Such dark patterns undermine the original intent of permission explanations, shifting from facilitating informed consent to subtly coercing user decisions, ultimately reducing transparency and weakening users’ control over their personal data.

\subsection{Motivation}
Prior work has explored various dark patterns that mobile app developers can use to mislead users into taking unintended actions~\cite{mohamed2024attention,Tales2016,di2020ui,Comparative2022}.
In the context of permission requests, however, less is known about whether and how real-world Android apps adopt any strategies within developer-defined explanations to influence users’ decisions. 
Key questions remain unresolved: what types of dark patterns appear in permission explanations, how can these practices be systematically identified, and what mechanisms can mitigate such misuse?
These knowledge gaps motivate our systematic investigation into current permission explanation practices and development of strategies for enhancing their transparency and integrity.

\section{Empirical Study}
To understand the current landscape of permission explanation practices and identify systematic deficiencies, we conducted an empirical study examining 600 mobile apps. This empirical study serves two purposes: establishing a scientifically grounded evaluation framework and providing empirical evidence of transparency gaps in real-world implementations.

\subsection{Hierarchical Compliance Review Framework}
To systematically evaluate permission explanation compliance and quality, we developed a three-layer hierarchical review framework. Since existing regulatory guidelines provide only high-level principles without specific evaluation criteria, we employed expert consensus building to operationalize these requirements into practical assessment dimensions.

\subsubsection{Expert Consensus Building}
We began by examining key regulatory documents including GDPR, PIPL, and industry guidelines such as the Chinese ``Guidelines for the Use of System Permissions in Mobile Internet Applications~\cite{tc260}.'' These regulations emphasize principles such as purpose clarity, user accessibility, and avoidance of misleading content, but lack specific operational guidance for systematic evaluation.

To establish a scientifically grounded and practically applicable evaluation framework, we adopted the Delphi method~\cite{delphi} to facilitate structured expert consultation. This approach is particularly suitable for areas lacking unified standards and enables systematic development of evaluation criteria through multiple rounds of anonymous expert feedback~\cite{Okoli2004Delphi,Varona_2021}.
We recruited six domain experts: two privacy researchers from academia, two mobile app developers from the technology industry, and two compliance professionals from regulatory agencies, all with over three years of relevant experience. The consultation process comprised two rounds:

\begin{itemize}[noitemsep, topsep=1pt, partopsep=1pt, listparindent=\parindent, leftmargin=*]
\item \textbf{Round 1:} Experts received comprehensive materials including summaries of regulatory principles and requirements from major privacy regulations. Without predefined evaluation dimensions, experts independently identified critical aspects that should be assessed when evaluating permission explanation compliance. Through open-ended questionnaires, experts proposed various evaluation dimensions including visibility, accessibility, clarity, specificity, comprehensibility, completeness, accuracy, third-party disclosure, purpose alignment, and avoidance of misleading content. We synthesized expert responses and identified recurring themes across these proposed dimensions.
\item \textbf{Round 2:} Based on Round 1 synthesis, we organized the identified dimensions into three hierarchical layers: explanation visibility (whether explanations are accessible to users), linguistic appropriateness (whether explanations are clearly and appropriately expressed), and substantive accuracy (whether explanations truthfully reflect actual permission usage). We presented this three-tier framework to experts. 
Kendall's W coefficient analysis showed values exceeding 0.75 for all framework components, indicating strong expert consensus and satisfying established Delphi standards~\cite{su132112019}.
\end{itemize}

\subsubsection{Three-Layer Evaluation Framework}
Drawing on expert feedback, we finalized a layered structure (as follows) for systematically evaluating permission explanations, which serves as the basis for our empirical analysis and tool development.

\begin{itemize}[noitemsep, topsep=1pt, partopsep=1pt, listparindent=\parindent, leftmargin=*]
\item \textbf{L1: Visual Accessibility.} This layer examines whether the explanation is visibly presented by the time of the permission request. The explanation must not be concealed in less accessible resources like privacy policies or other hidden information pages.
Only apps that demonstrate a visible explanation proceed to further evaluation stages.

\item \textbf{L2: Linguistic Appropriateness.} This layer evaluates the clarity, specificity, and appropriateness of the explanation text. The explanation should (1) explicitly state the purpose of the permission, avoiding vague or overly broad terms; (2) use plain and understandable language, avoiding obscure technical jargon; and (3) not contain any misleading claims that exaggerate necessity or pressure the user. This assessment primarily relies on semantic and textual analysis.

\item \textbf{L3: Substantive Accuracy.} This layer assesses whether the explanation truthfully reflects the actual use of the requested permission. It checks for completeness and honesty, ensuring that all relevant uses, including those by third-party SDKs or data-sharing practices, are clearly disclosed. This step needs to integrate dynamic analysis techniques to compare the app's real behavior with the explanation content and identify potential inconsistencies.
\end{itemize}

This three-layered framework establishes a progressive, in-depth compliance evaluation structure that spans accessibility, linguistic appropriateness, and behavioral alignment. Firmly grounded in legal requirements, practical insights, and expert consensus, it serves as the foundation for subsequent evaluation and system design in this study.

\subsection{State-quo of Permission Explanations}

With this three-layered framework, we conducted a systematic analysis of 600 real-world mobile apps across various categories from the Chinese marketplace, aiming to examine the implementation status of permission explanations in practice.

\subsubsection{Analysis Process}
Given the limitations of automated UI exploration tools in achieving comprehensive permission coverage, we adopted manual testing to ensure maximum triggering of permission requests. 
Trained researchers systematically explored each app for at least 5 minutes, focusing on triggering permission-sensitive features such as location services, camera access, contact management, and media access.  
During testing, we applied Frida~\cite{frida}, a dynamic instrumentation toolkit, to hook Android system APIs at runtime and capture call stack traces for sensitive resources, including calendars, call logs, contacts, location, device IDs, body sensors, SMS, storage, camera, and microphone.
We captured screenshots of all permission request dialogs and simultaneously monitored corresponding sensitive API calls.

Two authors with relevant expertise independently evaluated each collected permission explanation following the three-layer hierarchical framework: First, the explanation visibility (L1) was assessed. If this was deemed acceptable, the linguistic appropriateness (L2) was evaluated, and the substantive accuracy was examined using the captured API call information (L3). Any disagreements were resolved through consensus discussions with all authors.

\subsubsection{Key Findings}

Through systematic analysis of 600 apps, we captured a total of 1,031 permission requests across various sensitive resources. Our evaluation using the three-layer framework revealed significant deficiencies in current permission explanation practices:

\begin{itemize}[noitemsep, topsep=1pt, partopsep=1pt, listparindent=\parindent, leftmargin=*]
\item \textbf{Missing Explanations:} Among the 1,031 permission requests, 158 cases (15\%, involving 89 apps) provided no explanatory explanation whatsoever. These apps directly presented system permission dialogs without any contextual information about permission purposes, making it difficult for users to make informed decisions.

\item \textbf{Linguistic Appropriateness Issues:} Among the 873 explanations that were present, 70 cases (involving 60 apps) failed to meet linguistic appropriateness standards. The primary issues included:
\textit{Lack of specificity (43 cases):} explanations used vague language such as ``to provide better services'' without explaining specific use contexts.
\textit{Complexity and jargon (9 cases):} Technical terminology like ``SDK'' or ``LBS services'' that ordinary users might not understand.
\textit{Manipulative language (18 cases):} Misleading claims that exaggerated necessity or employed pressure tactics, such as ``must authorize for normal use'' without justification.

\item \textbf{Substantive Accuracy Concerns:} Most critically, 246 explanations (28\% of those with explanation, involving 159 apps) demonstrated incomplete information disclosure, particularly regarding third-party SDK usage. This represents a conservative estimate as many permission usages may not have been triggered during our testing process. Our API analysis revealed many cases where advertising or analytics SDKs accessed permissions without any mention in the accompanying explanations.
For example, one navigation app stated that location access was ``for route planning and nearby vehicle finding,'' but our analysis detected that a ByteDance advertising SDK (\texttt{com.bytedance.sdk.openadsdk}) was also accessing location data in background threads—a usage completely undisclosed to users.
\end{itemize}

Modern apps typically integrate 5-10 third-party SDKs (with some containing up to 50)~\cite{appicaptorblog}, and these components inherit all permissions granted to the host app. However, our analysis showed that explanations consistently focused solely on primary app functionality while omitting third-party usage scenarios.
This represents a transparency gap where users consent to specific stated purposes while remaining unaware of secondary data uses by integrated SDKs.

\subsection{Problem Statement}

Our empirical analysis reveals substantial deficiencies in current permission explanation practices that undermine their intended transparency benefits. Three critical problems emerge: (1) widespread absence of explanations (15\% of permission requests), (2) linguistic barriers preventing user comprehension due to vague language and technical jargon, and (3) systematic omission of third-party SDK data usage despite API calls by these components being clearly detected in our monitoring.

These deficiencies represent fundamental violations of informed consent principles mandated by privacy regulations. Users cannot make genuinely informed decisions when explanations are missing, unclear, or incomplete, while undisclosed third-party access creates privacy risks that users cannot anticipate or protect against.
The gap between regulatory intentions and practical implementation demonstrates the urgent need for systematic evaluation and optimization approaches that can automatically assess explanation quality and provide targeted improvement recommendations.

\section{\tool}

To address the identified challenges, we present \tool (\underline{\textbf{S}}mart \underline{\textbf{C}}ompliance and \underline{\textbf{O}}ptimization of \underline{\textbf{P}}ermission \underline{\textbf{E}}xplanations), an automated multi-agent framework that systematically evaluates permission explanation compliance and generates targeted optimization recommendations. 
Built upon OpenAI Agents SDK~\cite{agentsdk}, \tool embodies key characteristics of intelligent agent systems such as autonomous planning, reasoning, and action, 
enabling systematic assessment across the three-layer evaluation framework established in our empirical study.

\begin{figure}
    \centering
    \includegraphics[width=0.46\textwidth]{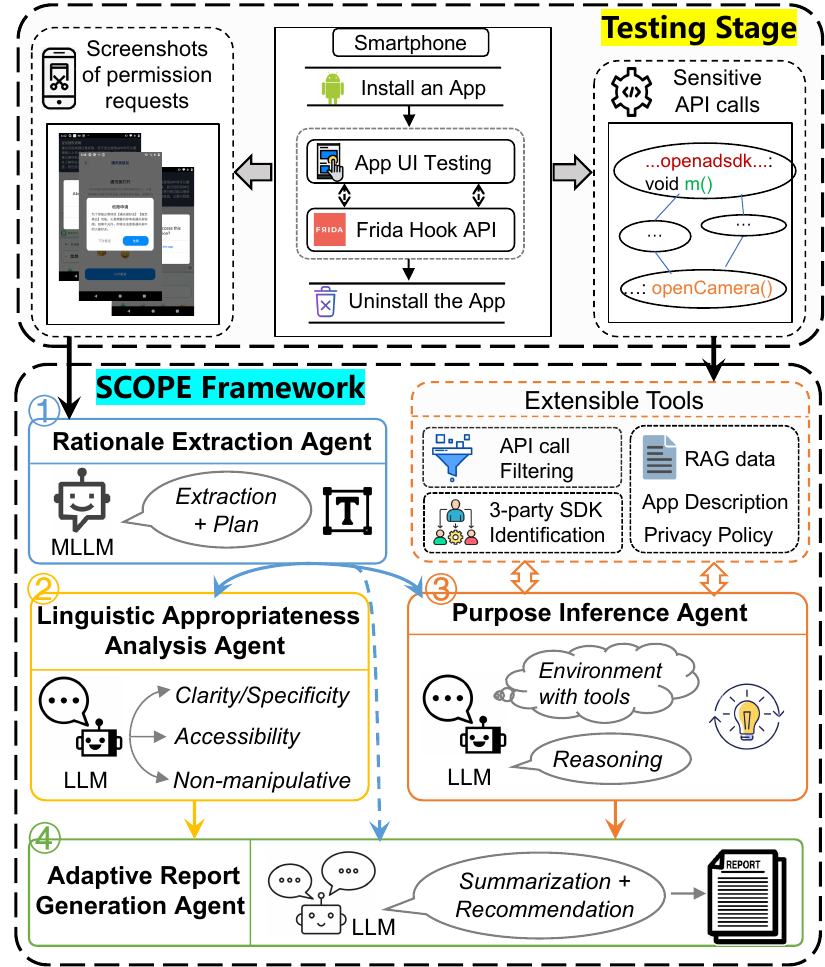}
    \vspace{-0.1in}
    \caption{Overview of the \tool workflow.}
    \vspace{-0.1in}
    \label{fig:ToolArchitecture}
\end{figure}

\subsection{Overall Architecture}

\tool adopts a collaborative multi-agent architecture where specialized agents work together through structured planning, execution, and communication mechanisms. 
As illustrated in \autoref{fig:ToolArchitecture}, the framework orchestrates four specialized agents 
and follows a structured pipeline with conditional branching based on intermediate results:

\textbf{Explanation Extraction Agent} processes captured screenshots to identify and extract permission explanation texts using multimodal large language models (MLLM). The agent employs planning capabilities to determine the execution path: if explanation text is successfully extracted, the coordinator routes the content to both Linguistic Appropriateness Analysis Agent and Purpose Inference Agent for parallel analysis; if no explanation is found, the workflow bypasses linguistic and purpose analysis, directing control to the Adaptive Report Generation Agent to document the absence finding.

\textbf{Linguistic Appropriateness Analysis Agent} evaluates extracted explanation texts for regulatory compliance using few-shot learning approaches. This agent maintains memory of linguistic patterns and compliance criteria, enabling consistent evaluation across different explanation styles.

\textbf{Purpose Inference Agent} analyzes runtime API call information and app metadata to determine actual permission usage scenarios. The agent employs Retrieval-Augmented Generation (RAG) to incorporate app descriptions and privacy policy information, combined with contextual reasoning and third-party SDKs identification, enabling comprehensive purpose inference and truthfulness assessment.

\textbf{Adaptive Report Generation Agent} synthesizes results from all preceding agents to generate comprehensive evaluation reports. This agent aggregates findings, identifies compliance gaps, and generates targeted recommendations based on the specific combination of issues detected.

\subsection{Explanation Extraction Agent}

The explanation extraction agent processes captured screenshots to identify and extract permission explanation texts. Given the diverse presentation formats of permission explanations across different apps, this agent must accommodate various UI styles, text layouts, and visual designs.

We considered multiple text extraction approaches, including traditional OCR techniques (e.g., Tesseract, PaddleOCR) and multimodal large language models (e.g., gemini 2.0 flash, gpt-4o). While traditional OCR methods provide basic text recognition capabilities, they face significant limitations when dealing with complex UI scenarios, including low contrast text, complex UI elements, and varying font styles. Moreover, OCR methods extract raw text without contextual understanding, requiring additional post-processing to filter relevant permission-related content.
In contrast, MLLMs demonstrate superior performance by directly understanding interface context and leveraging their language modeling capabilities to pinpoint permission explanation text. Our preliminary evaluation showed that MLLMs achieve higher accuracy in text extraction, better adaptability to diverse UI designs, and more efficient end-to-end processing by eliminating the need for extensive post-processing.
We thus adopt MLLMs as the core technology for explanation extraction.

The agent employs carefully designed prompts to guide the MLLM in precisely identifying and extracting explanation content. As shown below, the prompt includes specific instructions for recognizing permission-related explanations while filtering out irrelevant interface elements.

\begin{tcolorbox}[colback=gray!10, colframe=gray!80, title={\small\textbf{Prompt Example 1: explanation Extraction}}, fonttitle=\bfseries, left=3pt, right=3pt]
\small
\textbf{System:} You are an assistant who helps extract text from images.

\textbf{User:} Analyze this mobile app screenshot and extract any explanatory text that justifies permission requests. Focus on developer-provided explanations that clarify why specific permissions are needed, not system-generated permission dialogs.

Extract the exact explanation text if present. If no developer explanation exists, respond with ``No explanation found.''

\textbf{Output format:}
\begin{itemize}
\item Permission Type: [specific permission being requested]
\item Explanation Text: [exact explanation text for the permission]
\end{itemize}
\end{tcolorbox}

\subsection{Linguistic Appropriateness Analysis Agent}

This agent evaluates the quality of extracted explanation texts to ensure they meet regulatory requirements for clarity, specificity, and non-manipulative communication, 
three critical dimensions established through expert consensus in our empirical study.
To enhance the accuracy and adaptability of text quality assessment, this agent employs few-shot learning techniques with LLMs. Unlike traditional rule-based approaches that rely on predefined keyword libraries or readability metrics, the LLM-based method provides more nuanced semantic understanding and better generalization across diverse expression styles.
The agent evaluates explanation texts across the following three key dimensions:

\textbf{(1) Clarity and Specificity:} The agent assesses whether the explanation provides concrete, specific explanations of permission usage rather than vague, generic statements. It identifies ambiguous terms and evaluates the overall comprehensibility of the explanation.

\textbf{(2) Language Accessibility:} The analysis examines whether the language is accessible to average users, avoiding excessive technical jargon or complex legal terminology that may impede user understanding.

\textbf{(3) Non-manipulative Communication:} The agent detects potentially manipulative or coercive language patterns, including pressure tactics, emphasis on benefits while downplaying privacy implications, or use of persuasive techniques designed to unduly influence user decisions.

The few-shot learning approach incorporates multiple annotated examples in the prompt to guide the LLM in making accurate assessments, as shown in the following example.

\begin{tcolorbox}[colback=gray!10, colframe=gray!80, title={\small\textbf{Prompt Example 2: Linguistic Appropriateness Analysis}}, fonttitle=\bfseries, left=3pt, right=3pt]
\small
\textbf{System:} You are an expert in text semantics and linguistics.

\textbf{User:} Evaluate the following permission explanation text across three dimensions:

\textbf{Evaluation Criteria:}
\begin{enumerate}
\item \textbf{Clarity \& Specificity:} Is the purpose concrete and specific?
\item \textbf{Language Accessibility:} Is it understandable without technical expertise?
\item \textbf{Non-manipulative Communication:} Does it avoid coercive or unduly persuasive language?
\end{enumerate}

\textbf{Examples:}
\begin{itemize}
\item \textbf{Clarity \& Specificity:}
  \begin{itemize}
  \item \textit{Compliant:} ``We access your photo album to upload profile pictures.''
  \item \textit{Non-compliant:} ``Photo album permission needed to optimize your experience.'' (too vague)
  \end{itemize}
\item \textbf{Language Accessibility:}
  \begin{itemize}
  \item \textit{Compliant:} ``We use location data to show nearby stores.''
  \item \textit{Non-compliant:} ``Location-based services (LBS) provide POI information.'' (too technical)
  \end{itemize}
\item \textbf{Non-manipulative Communication:}
  \begin{itemize}
  \item \textit{Compliant:} ``We access your contact list for the contact synchronization feature.''
  \item \textit{Non-compliant:} ``Allowing contacts permission unlocks premium services.'' (manipulative incentive)
  \end{itemize}
\end{itemize}

\textbf{explanation to evaluate:} "[EXPLANATION\_TEXT]"

\textbf{Response format:}
\begin{itemize}
\item Overall Compliance: [Compliant/Non-compliant]
\item Issues Identified: [list specific problems]
\end{itemize}
\end{tcolorbox}

\subsection{Purpose Inference Agent}

The purpose inference agent determines the actual usage scenarios of requested permissions, enabling verification of explanation truthfulness and completeness. Prior studies have demonstrated that extracting contextual cues from sensitive API calls and app metadata (e.g., app descriptions, privacy policies) can effectively infer an app's actual purpose for data access~\cite{yang2022describectx,11023406}. Inspired by these findings, our agent 
orchestrates multiple analytical components through a systematic workflow.
First, the agent invokes the API Call Filtering Tool to obtain temporally relevant API traces associated with each permission request. Simultaneously, it calls the Third-party SDK Identification Tool to determine the attribution of each API call, distinguishing between main app functionality and third-party SDK usage. With this foundational information, the agent employs RAG-enhanced context integration to retrieve relevant app metadata from app descriptions and privacy policies. By analyzing naming patterns, execution contexts, and semantic cues from multiple sources, the agent infers the actual purposes of permission usage and compares them against the declared explanations, identifying discrepancies in both completeness (e.g., undisclosed third-party usage) and accuracy (e.g., stated purposes inconsistent with API usage). This workflow enables systematic detection of explanation truthfulness violations and provides detailed insights for improving permission transparency.

\subsubsection{API Call Filtering Tool}

This tool extracts relevant API call traces from comprehensive runtime monitoring data captured during dynamic analysis. The tool operates using temporal correlation analysis: when a permission request occurs, it defines a time window around the request timestamp and captures all API calls within this window. The tool then performs deduplication to remove redundant traces and filters out irrelevant system calls, retaining only permission-sensitive API invocations that are temporally associated with the permission request. This filtered set of API calls provides the foundational data for subsequent purpose inference analysis.

\subsubsection{Third-party SDK Identification Tool}

This tool identifies whether permission usage originates from the main app or integrated third-party SDKs by analyzing calling package names and determining their attribution.
The tool operates using a hybrid approach combining a reference database with LLM-based inference:
(1) \textit{Reference Database:} We initially compiled a comprehensive database of known third-party SDKs from multiple sources, including Google's official documents, open-source academic artifacts, and industry reports. This database maps package names to their corresponding service categories (e.g., advertising, analytics, social features).
(2) \textit{LLM-Enhanced Expansion:} Recognizing that the reference database cannot be exhaustive, we integrate an LLM component to handle unknown package names. When a package name is not found in the database, the tool queries the LLM to determine whether the package likely represents a third-party SDK based on naming patterns, semantic cues, and the LLM's inherent knowledge. Validated predictions are then incorporated back into the database, enabling continuous refinement and expansion of our SDK identification capabilities.

\subsubsection{RAG-Enhanced Context Integration}

To enhance purpose inference accuracy, the agent employs RAG to incorporate relevant app metadata alongside API call analysis. The agent maintains a knowledge base containing app store descriptions and privacy policy segments, enabling contextual enrichment when API call semantics are insufficient.
When analyzing permission usage, the agent first retrieves relevant context from the app's description to understand its primary functionalities, then searches the privacy policy for existing permission usage statements. This contextual information is integrated with API call analysis to provide comprehensive purpose inference.

\begin{tcolorbox}[colback=gray!10, colframe=gray!80, title={\small\textbf{Prompt Example 3: Purpose Inference}}, fonttitle=\bfseries, left=3pt, right=3pt]
\small
\textbf{System:} You are an Android security analysis expert.

\textbf{User:} Analyze the following API call information, app context, and privacy policy statement to infer the actual purpose of permission usage:

\textbf{API Call Information:}
\begin{itemize}
\item Permission: Location Access
\item API Method: \texttt{getLastKnownLocation()}
\item Calling Class: \texttt{LocationTracker}
\item Call Stack: \texttt{com.example...AdNetwork $\rightarrow$ ... $\rightarrow$ android.location...getLastKnownLocation}
\item Caller Type: Third-party SDK (Advertising)
\end{itemize}

\textbf{App Context:} The app provides route planning and traffic updates ...

\textbf{Privacy Policy Statement:} ``We use location data to provide navigation services and route planning.''

\textbf{Task:} 
\begin{enumerate}
\item Infer the actual purpose of this API call based on the provided context
\item Identify the entity responsible for the permission usage
\item Provide evidence supporting your inference
\end{enumerate}

\textbf{Response format:}
\begin{itemize}
\item Inferred Purpose: [actual usage scenario]
\item Caller Entity: [main app/third-party SDK]
\item Evidence: [key indicators from API calls and context that support the inference]
\end{itemize}
\end{tcolorbox}

\subsection{Adaptive Report Generation Agent}

The final agent synthesizes analysis results from all preceding agents to generate comprehensive evaluation reports with targeted optimization recommendations. This agent serves as the integration point for the multi-agent framework, providing actionable insights for developers to improve their permission explanation practices.

\subsubsection{Compliance Assessment}

The agent evaluates the overall compliance status based on the prior three agents:
\begin{itemize}[noitemsep, topsep=1pt, partopsep=1pt, listparindent=\parindent, leftmargin=*]
\item \textbf{Visibility Compliance:} Whether explanations are present and accessible to users during permission requests;
\item \textbf{Linguistic Compliance:} Whether explanation texts meet standards for clarity, specificity, and non-manipulative;
\item \textbf{Truthfulness Compliance:} Whether stated purposes accurately reflect actual permission usage, including disclosure of third-party data access.
\end{itemize}

\subsubsection{Targeted Recommendation Generation}

Based on identified compliance gaps, the agent generates specific recommendations tailored to each app's particular issues. Recommendations are categorized by compliance layer and include:

\begin{itemize}[noitemsep, topsep=1pt, partopsep=1pt, listparindent=\parindent, leftmargin=*]
\item \textbf{Content Improvements:} Specific suggestions for enhancing explanation text clarity, replacing ambiguous terms, and providing more concrete usage explanations;
\item \textbf{Disclosure Enhancements:} Recommendations for addressing incomplete disclosure of third-party SDK data usage, including suggested language for transparent communication about data sharing;
\end{itemize}

The adaptive nature of this agent ensures that recommendations are contextually relevant and practically implementable, supporting developers in creating more transparent and regulation-compliant permission practices that ultimately enhance user trust and informed consent.

\begin{tcolorbox}[colback=gray!10, colframe=gray!80, title={\small\textbf{Prompt Example 4: Adaptive Report Generation}}, fonttitle=\bfseries, left=3pt, right=3pt]
\small
\textbf{System:} You are a compliance analysis expert.

\textbf{User:} Generate a compliance evaluation report based on the following analysis results:

\textbf{Analysis Results:}
\begin{itemize}
\item \textbf{Explanation Extraction Result:} [EXTRACTION\_RESULT]
\item \textbf{Linguistic Analysis Result:} [LINGUISTIC\_ANALYSIS]
\item \textbf{Purpose Inference Result:} [PURPOSE\_INFERENCE]
\end{itemize}

\textbf{App Information:}
\begin{itemize}
\item App Name: [APP\_NAME]
\item Permission Type: [PERMISSION\_TYPE]
\end{itemize}

\textbf{Task:}
\begin{enumerate}
\item Evaluate overall compliance status across the three-layer framework
\item Generate targeted recommendations for each identified issue
\item Provide implementation guidance and priority ranking
\end{enumerate}

\textbf{Response format:}
\begin{itemize}
\item \textbf{Compliance Summary:} [status for visibility, linguistic, and truthfulness layers]
\item \textbf{Recommendations:} [actionable improvements categorized by compliance layer]
\item \textbf{Implementation Priority:} [ranked list of suggested improvements]
\end{itemize}
\end{tcolorbox}
\section{Evaluation}

To assess the effectiveness of our proposed multi-agent framework for evaluating and optimizing permission explanations, we conducted comprehensive experiments using the dataset of 600 real-world mobile apps from our empirical study. These apps had been manually annotated by domain experts, providing ground truth for evaluation. Our experiments examine the accuracy and performance of each agent within the framework, as well as the practical utility of the generated optimization recommendations.

\begin{itemize}
\item\textbf{RQ1:} How effective is the explanation extraction agent in identifying and extracting permission explanation texts from app screenshots?
\item \textbf{RQ2:} How accurate is the linguistic appropriateness agent in semantic analysis and compliance assessment?
\item \textbf{RQ3:} How precise is the purpose inference agent in determining actual permission usage purposes?
\item \textbf{RQ4:} Do the optimization recommendations generated by our framework positively impact user privacy perception and understanding?
\end{itemize}

\subsection{RQ1: Explanation Extraction Effectiveness}
\subsubsection{Experimental Setup}

We evaluated the explanation extraction agent's capability to automatically identify and extract permission explanation texts from app interfaces. The experiment compared traditional OCR tools (Tesseract~\cite{tesseract}) against several popular MLLMs (gpt-4-vision-preview, gpt-4o, gemini 2.5 flash, claude-3-sonnet).

We evaluated extraction results across 100 permission request screenshots collected from our dataset. The evaluation measured two key dimensions: (1) permission type identification using Permission Matching Rate (PMR), which indicates the proportion of correctly identified permission types; (2) explanation text extraction using Usage Extraction Accuracy (UEA), which indicates the proportion of accurately extracted explanation texts that completely match human annotations.

\subsubsection{Results}

\autoref{tab:extraction_results} summarizes the experimental results across different methods.
For permission identification, all methods achieved nearly perfect performance with PMR scores approaching 100\%, indicating that identifying permission types from app screenshots is a relatively straightforward task across different approaches. However, substantial performance disparities emerged in explanation text extraction. Tesseract OCR exhibited significant extraction limitations with only 50\% UEA, particularly struggling with dark backgrounds and complex UI layouts. Conversely, MLLMs achieved considerably higher extraction accuracy ranging from 65\% to 98\%.
Among the MLLMs, gpt-4-vision-preview achieved the highest performance with 98\% UEA, followed by gpt-4o (92\%), gemini 2.5 flash (82\%), and claude-3-sonnet (65\%). The gpt-4 variants demonstrated superior performance in handling complex UI layouts, diverse text formats, and low-contrast scenarios compared to other models. 
Besides, a critical advantage of MLLMs over traditional OCR is their ability to directly identify permission-relevant explanatory content by integrating visual context with semantic understanding, whereas OCR indiscriminately extracts all text and requires extensive post-processing to isolate relevance.

\vspace{-0.1in}
\begin{table}[htbp]
\centering
\caption{Performance of Explanation Text Extraction Methods}
\vspace{-0.1in}
\label{tab:extraction_results}
\begin{tabular}{l|c|c|p{3cm}}
\hline
\textbf{Method} & \textbf{PMR} & \textbf{UEA} & \textbf{Notes} \\
\hline
Tesseract OCR & 98\% & 50\% & Struggle with dark backgrounds, complex layouts \\
gpt-4-vision-preview & 100\% & 98\% & Excellent across various UI designs \\
gpt-4o & 100\% & 92\% & Strong performance, minor variations \\
gemini 2.5 flash & 100\% & 82\% & Good overall, occasional context misunderstanding \\
claude-3-sonnet & 100\% & 65\% & Struggle with complex scenarios \\
\hline
\end{tabular}
\vspace{-0.1in}
\end{table}

\begin{framed}
\vspace{-0.05in}
\noindent\textbf{Answer to RQ1:} The explanation extraction agent demonstrates high effectiveness, with MLLMs achieving up to 98\% extraction accuracy. MLLMs exhibit superior performance across diverse UI designs and can directly identify permission-relevant content without extensive post-processing, outperforming traditional OCR.
\vspace{-0.05in}
\end{framed}

\subsection{RQ2: Linguistic Appropriateness Analysis Accuracy}

\subsubsection{Evaluation Setup}

This experiment evaluated the linguistic appropriateness analysis agent's performance in semantic understanding and compliance violation detection. We compared our proposed LLM-based few-shot learning approach against traditional keyword matching and rule-based methods commonly used in existing research.

The keyword matching and rule-based baseline employed the following detection rules: vague expressions (e.g., ``may,'' ``some,'' ``related''), coercive or threatening language (e.g., ``must,'' ``otherwise,'' ``mandatory''), and technical terminology (e.g., ``SDK,'' ``IMEI,'' ``MAC address''). We enhanced this baseline with structured semantic rules including sentence segmentation strategies and negation context analysis to improve accuracy and reduce misclassifications. This comprehensive comparison enables systematic evaluation of different approaches to linguistic appropriateness assessment.

We tested multiple mainstream LLMs (gpt-4o, claude-3-sonnet, gemini-2.5-pro, grok-3, deepseek-v3, doubao-1.5-pro-256k, qwen3-32b) using few-shot learning prompts containing multiple examples with corresponding labels. The evaluation dataset comprised 100 permission explanation texts from our empirical study, manually annotated as ``compliant" (50 texts) or ``non-compliant" (50 texts) with specific violation types.
We measured three key metrics: (1) Evaluation Consistency (EC): proportion of model judgments matching expert annotations; (2) False Positive Rate (FPR): proportion of compliant texts incorrectly flagged as violations; and (3) False Negative Rate (FNR): proportion of actual violations missed by the system.

\subsubsection{Results}

\autoref{tab:linguistic_results} presents the performance comparison across different methods.
The results reveal a substantial performance gap between LLM-based few-shot learning and traditional rule-based approaches. The keyword matching baseline demonstrated poor overall effectiveness with only 57.5\% evaluation consistency, suffering from both high FNR (70\%) and FPR (33.3\%). This indicates fundamental limitations of rule-based methods in capturing the semantic nuances of linguistic appropriateness violations, as they rely on surface-level pattern matching rather than contextual understanding.

Among LLMs, qwen3-32b achieved the highest performance (93.5\% EC) with balanced error rates (8.5\% FPR, 4\% FNR). Models exhibited distinct error patterns: gpt-4o demonstrated perfect recall (0\% FNR) but higher false positives (23\%), while gemini-2.5-pro and deepseek-v3 achieved perfect precision (0\% FPR) but higher false negatives (30-35\%).
A notable finding is the systematic trade-off pattern observed across models: those achieving perfect precision (0\% FPR) consistently exhibited higher false negative rates, while models with perfect recall (0\% FNR) showed elevated false positive rates. This pattern reflects different decision boundaries that models establish when processing borderline cases. The task-specific calibration of decision boundaries could potentially improve model performance for linguistic appropriateness assessment by optimizing the precision-recall balance according to specific app requirements.

\begin{table}[htbp]
\centering
\vspace{-0.1in}
\caption{Performance of Linguistic Appropriateness Analysis Methods}
\vspace{-0.1in}
\label{tab:linguistic_results}
\begin{tabular}{l|c|c|c}
\hline
\textbf{Method} & \textbf{EC (\%)} & \textbf{FPR (\%)} & \textbf{FNR (\%)} \\
\hline
Keyword Matching \& Rules & 57.5 & 33.3 & 70 \\
qwen3-32b & 93.5 & 8.5 & 4 \\
gpt-4o & 85 & 23 & 0 \\
gemini-2.5-pro & 85 & 0 & 30 \\
deepseek-v3 & 82.5 & 0 & 35 \\
doubao-1.5-pro-256k & 82.5 & 12 & 25 \\
grok-3 & 70 &25 & 27.8 \\
claude-3-sonnet & 67.5 & 23 & 50 \\
\hline
\end{tabular}
\end{table}

\begin{framed}
\vspace{-0.05in}
\noindent\textbf{Answer to RQ2:} The linguistic appropriateness analysis agent demonstrates high effectiveness, with qwen3-32b achieving 93.5\% evaluation consistency and gpt-4o reaching 85\%. LLM-based few-shot learning significantly outperforms traditional keyword matching, with models exhibiting distinct conservative vs. aggressive assessment strategies depending on their precision-recall trade-offs.
\vspace{-0.05in}
\end{framed}

\subsection{RQ3: Purpose Inference Accuracy}

\subsubsection{Experimental Setup}

This experiment assessed the purpose inference agent's accuracy in identifying apps' actual permission usage intentions by analyzing API call stack information combined with app context.
We selected 50 permission usage cases from our dataset, covering different permission types (location, contacts, storage) and usage scenarios. Each case was manually labeled by domain experts with actual usage purposes and calling entities (main app or third-party SDKs). We used gpt-4o as the base inference model, combining call stack context with app background information.

We conducted two groups of experiments to evaluate the inference performance with different inputs: (1) Control Group: provided only API call stack information, and (2) Experimental Group: provided API call information plus metadata including app store descriptions and privacy policies.
We measured Usage Inference Accuracy (UIA) as the proportion of correctly inferred permission purposes.

\subsubsection{Results}

The results demonstrate the effectiveness of our approach for permission purpose inference. The experimental group achieved 92\% UIA (46/50), while the control group achieved 76\% (38/50), suggesting that while API call stack information alone can provide reasonable capability for purpose identification, the addition of app contextual information significantly enhances inference accuracy.
This improvement is particularly valuable when API call stacks provide limited semantic cues or when class/method names are ambiguous. For example, in one case where an app called \texttt{getLastKnownLocation()} from a generic \texttt{BaseService.doTask()} method, the model achieved reasonable inference from the call stack alone but successfully pinpointed the specific purpose when provided with the app's store description (``provides urban route planning and real-time transit navigation'').
These results validate the feasibility of API-based purpose inference and the effectiveness of our context-enhanced reasoning approach for achieving high-precision permission purpose identification.

\begin{framed}
\vspace{-0.05in}
\noindent\textbf{Answer to RQ3:} The purpose inference agent demonstrates high effectiveness, with API call analysis achieving 76\% accuracy and context-enhanced reasoning reaching 92\%. The approach successfully enables automated purpose identification, with app contextual information providing significant additional improvement.
\vspace{-0.05in}
\end{framed}

\subsection{RQ4: User Study on Optimization Effectiveness}

\subsubsection{Experimental Setup}

To evaluate whether our framework's optimization recommendations positively impact user privacy perception, we conducted a user study with 30 participants comparing original versus optimized permission explanations.
Participants were presented with 5 pairs of permission explanations (original and optimized versions) and asked to select which version they found clearer and more understandable. We used a within-subjects design where each participant evaluated all pairs, resulting in 150 total comparisons. The study included both quantitative preference measurements and qualitative feedback through open-ended questions about participants' selection explanations.

\subsubsection{Results}

Results showed that 127 out of 150 comparisons (84.6\%) favored the optimized explanation versions over the original versions. At the individual level, 18 participants (60\%) chose optimized versions for all 5 tasks, while 27 participants (90\%) selected optimized versions for more than half their tasks.
Qualitative feedback supported these quantitative findings. Most participants noted that optimized explanations were more specific in expression, used simpler language, reduced technical jargon and vague statements, and better facilitated understanding of permission purposes and privacy risk identification. Some users also mentioned that optimized versions increased their trust in apps and made them more willing to make informed authorization decisions.
These results validate the effectiveness of our intelligent evaluation and optimization framework in enhancing user privacy perception and understanding.

\begin{framed}
\vspace{-0.05in}
\noindent\textbf{Answer to RQ4:} The optimization recommendations significantly improve user privacy perception, with 84.6\% of users preferring optimized explanations over originals. Users found optimized versions clearer, more specific, and more trustworthy.
\vspace{-0.05in}
\end{framed}

\section{Discussion}

\subsection{Extensibility}

While our study used manual testing for comprehensive coverage, the framework seamlessly integrates with automated UI exploration tools like DroidBot~\cite{droidbot} and Fastbot~\cite{fastbot} by requiring only permission screenshots and API traces as input, making it agnostic to data collection methods.
The modular architecture enables independent scaling and customization of agents to meet specific organizational needs. \tool can be integrated into CI/CD pipelines for automatic explanation assessment during development
in enterprise environments.

\subsection{Implications}

\subsubsection{For App Developers} 
\tool provides developers with automated assessment and targeted recommendations, enabling evidence-based improvements in explanation clarity and regulatory compliance. The framework's detection of undisclosed third-party SDK usage is particularly valuable, helping developers identify hidden compliance violations. Integration into development workflows enables early detection of transparency issues, reducing post-deployment risks and supporting continuous compliance in agile development processes.

\subsubsection{For Privacy Regulators}
Our research provides the first systematic empirical evidence of inadequate permission explanation practices. \tool demonstrates feasible automated compliance assessment, offering scalable monitoring capabilities for large app portfolios. The three-layer evaluation framework provides a structured foundation for standardized compliance testing and future regulatory guidelines. Our findings on systematic third-party SDK disclosure failures highlight critical gaps requiring more specific regulatory requirements.

\subsection{Limitations and Future Work}

\subsubsection{Sample Representativeness and Subjective Annotation}
Our analysis of 600 apps from the Chinese market represents a limited sample of the broader mobile ecosystem.
Future work should expand to include long-tail apps, region-specific apps, and diverse developer categories.
Moreover, all annotation and evaluation tasks were performed by the author team, which, although experienced, may introduce unintentional subjective bias due to the lack of external reviewers. While inter-author consensus was used to resolve disagreements, this limitation should be addressed in future studies by involving a more diverse set of annotators, including external experts or crowd workers, to improve result validity and generalizability.

\subsubsection{Obfuscation and Inference Challenges}
Our purpose inference relies on semantic analysis of API calls, which is vulnerable to code obfuscation techniques that replace meaningful method names with meaningless strings. Future work should explore behavioral analysis techniques focusing on API call patterns, data flow analysis, surrounding code functions, and secondary cues like comments, to better interpret permission purposes. Dynamic loading and reflection-based API access also require more sophisticated analysis techniques.

\subsubsection{Framework Optimization and Scalability}
Components such as Linguistic Appropriateness and Compliance Assessment involve nuanced, context-dependent criteria that are hard to fully operationalize. Future work could mitigate this by applying self-consistency or multi-step reasoning, or by using ensembles across multiple models to improve stability and reduce variance.
Our current evaluation results may be influenced by specific prompt designs and the choice of LLMs, which could affect the framework's practical effectiveness. 
Considering economic constraints in practical deployment, future work will explore smaller open-source models and domain-specific fine-tuned models tailored to permission explanation analysis tasks. 

\section{Related Work}

\noindent\textbf{Data Transparency Mechanisms for Mobile Apps.}
Data transparency is critical for building user trust and enabling informed decision-making. Multiple stakeholders have promoted various privacy tools beyond traditional privacy policies, including Apple's Privacy Nutrition Labels~\cite{privacylabel}, App Tracking Transparency~\cite{att}, App Privacy Report~\cite{APR}, Google's Data Safety section~\cite{datasafety}, privacy dashboard~\cite{dashboard}, and permission explanations. However, growing research has highlighted significant limitations of these tools in supporting informed privacy decisions~\cite{zhang2022usable,koch2022keeping,degiulioAskAppNota,mohamed2024attention,11023406,wu2025transparency,khandelwal2024unpacking,baalous2025detecting}. Studies show that existing mechanisms often fail to provide actionable information, suffer from poor implementation practices, or lack systematic evaluation. Our work complements these transparency mechanisms by revealing systematic deficiencies in permission explanations and providing the first comprehensive framework to assess their real-world implementation quality.

\noindent\textbf{Dark Patterns in Permission Requests.}
Prior studies have identified manipulative design strategies, known as ``dark patterns,'' in permission interfaces. Gray et al.~\cite{gray2018dark} categorized dark patterns into nagging, obstruction, and forced action, which manifest through repetitive modals or concealed opt-out options. Other examples include feature tying, where core functionality is restricted unless permissions are granted~\cite{di2020ui}, and privacy zuckering, where consent information is obscured~\cite{Tales2016}. Most relevant is Mohamed et al.~\cite{mohamed2024attention}, who found that real-world permission prompts employed misleading, vague, or incentivizing language that confused users and distorted their understanding of permission implications. Our work extends this line of dark pattern research to permission explanations specifically, developing an automated framework that can systematically identify manipulative language patterns and provide targeted recommendations to mitigate such deceptive practices.
\section{Conclusion}

Our empirical analysis of 600 mainstream mobile apps reveals widespread deficiencies in permission explanation practices, including missing explanations, linguistic opacity, and systematic omission of third-party SDK data access. These failures violate regulatory transparency requirements and compromise informed consent.
We present \tool, a multi-agent framework for automated explanation evaluation and optimization that integrates MLLMs, few-shot learning, reasoning, and planning. Our evaluation demonstrates good performance of individual agents. 
User studies show 84.6\% preference for \tool-optimized explanations, validating practical utility.
\tool bridges the gap between regulatory intentions and practical implementation, offering a scalable solution for enhancing mobile application privacy transparency.

\section*{Acknowledgments}
This work was supported in part by the National Natural Science Foundation of China (grant No.62572209, 62172049) and the Hubei Provincial Key Research and Development Program (grant No. 2025BAB057).

\bibliographystyle{IEEEtran}
\bibliography{cite}

% Generated by IEEEtran.bst, version: 1.14 (2015/08/26)
\begin{thebibliography}{10}
\providecommand{\url}[1]{#1}
\csname url@samestyle\endcsname
\providecommand{\newblock}{\relax}
\providecommand{\bibinfo}[2]{#2}
\providecommand{\BIBentrySTDinterwordspacing}{\spaceskip=0pt\relax}
\providecommand{\BIBentryALTinterwordstretchfactor}{4}
\providecommand{\BIBentryALTinterwordspacing}{\spaceskip=\fontdimen2\font plus
\BIBentryALTinterwordstretchfactor\fontdimen3\font minus \fontdimen4\font\relax}
\providecommand{\BIBforeignlanguage}[2]{{%
\expandafter\ifx\csname l@#1\endcsname\relax
\typeout{** WARNING: IEEEtran.bst: No hyphenation pattern has been}%
\typeout{** loaded for the language `#1'. Using the pattern for}%
\typeout{** the default language instead.}%
\else
\language=\csname l@#1\endcsname
\fi
#2}}
\providecommand{\BIBdecl}{\relax}
\BIBdecl

\bibitem{cisareport}
CISA, ``Privacy and mobile device apps,'' \url{https://www.cisa.gov/news-events/news/privacy-and-mobile-device-apps}, 2022.

\bibitem{termsfeed}
R.~Bateman, ``Android collection of data and sensitive data,'' \url{https://www.termsfeed.com/blog/android-sensitive-data-collection/}, 2025.

\bibitem{androiddev2}
A.~Developers, ``Permissions on android,'' \url{https://developer.android.com/guide/topics/permissions/overview}, 2025.

\bibitem{betanews}
``Greedy apps collect more information than they should,'' \url{https://betanews.com/2023/10/27/greedy-apps-collect-more-information-than-they-should/}, 2023.

\bibitem{cybernews}
``Android financial apps too greedy for permissions,'' \url{https://cybernews.com/security/android-financial-apps-greedy-for-permissions/}, 2023.

\bibitem{bongard2022necessary}
K.~Bongard-Blanchy, J.-L. Sterckx, A.~Rossi, V.~Distler, S.~Rivas, and V.~Koenig, ``An (un) necessary evil-users'(un) certainty about smartphone app permissions and implications for privacy engineering,'' in \emph{2022 IEEE European Symposium on Security and Privacy Workshops (EuroS\&PW)}.\hskip 1em plus 0.5em minus 0.4em\relax IEEE, 2022, pp. 01--08.

\bibitem{190982}
\BIBentryALTinterwordspacing
P.~Wijesekera, A.~Baokar, A.~Hosseini, S.~Egelman, D.~Wagner, and K.~Beznosov, ``Android permissions remystified: A field study on contextual integrity,'' in \emph{24th USENIX Security Symposium (USENIX Security 15)}.\hskip 1em plus 0.5em minus 0.4em\relax Washington, D.C.: USENIX Association, 2015, pp. 499--514. [Online]. Available: \url{https://www.usenix.org/conference/usenixsecurity15/technical-sessions/presentation/wijesekera}
\BIBentrySTDinterwordspacing

\bibitem{appicaptorblog}
``Third-party library permission piggybacking in android apps,'' \url{https://blog.appicaptor.com/2025/02/27/third-party-library-permission-piggybacking-in-android-apps/}, 2025.

\bibitem{Heid_Sonntag_Heider_2025}
\BIBentryALTinterwordspacing
K.~Heid, E.~J. Sonntag, and J.~Heider, ``\BIBforeignlanguage{en}{Revisiting permission piggybacking of third-party libraries in android apps},'' 2025. [Online]. Available: \url{https://publica.fraunhofer.de/handle/publica/486121}
\BIBentrySTDinterwordspacing

\bibitem{GDPR}
E.~Commission, ``General data protection regulation,'' \url{https://gdpr.eu/tag/gdpr/}, 2025.

\bibitem{CCPA}
OAG, ``California consumer privacy act,'' \url{https://oag.ca.gov/privacy/ccpa}, 2025.

\bibitem{PIPL}
DigiChina, ``Personal information protection law of the people’s republic of china,'' \url{https://digichina.stanford.edu/work/translation-personal-information-protection-law-of-the-peoples-republic-of-china-effective-nov-1-2021/}, 2025.

\bibitem{pewresearch}
P.~R. Center, ``Americans’ attitudes and experiences with privacy policies and laws,'' \url{https://www.pewresearch.org/internet/2019/11/15/americans-attitudes-and-experiences-with-privacy-policies-and-laws/}, 2019.

\bibitem{syrenis}
Syrenis, ``Privacy policies: Is anyone reading them?'' \url{https://syrenis.com/resources/blog/privacy-policies-is-anyone-reading-them/}, 2024.

\bibitem{274679}
\BIBentryALTinterwordspacing
W.~Cao, C.~Xia, S.~T. Peddinti, D.~Lie, N.~Taft, and L.~M. Austin, ``A large scale study of user behavior, expectations and engagement with android permissions,'' in \emph{30th USENIX Security Symposium (USENIX Security 21)}.\hskip 1em plus 0.5em minus 0.4em\relax USENIX Association, 2021, pp. 803--820. [Online]. Available: \url{https://www.usenix.org/conference/usenixsecurity21/presentation/cao-weicheng}
\BIBentrySTDinterwordspacing

\bibitem{nngroup}
M.~Rosala, ``3 design considerations for effective mobile-app permission requests,'' \url{https://www.nngroup.com/articles/permission-requests/}, 2019.

\bibitem{mohamed2024attention}
R.~Mohamed, A.~Arunasalam, H.~Farrukh, J.~Tong, A.~Bianchi, and Z.~B. Celik, ``Attention please! an investigation of the app tracking transparency permission,'' in \emph{33rd USENIX Security Symposium (USENIX Security 2024)}, 2024, pp. 5017--5034.

\bibitem{hoober2019mobile}
S.~Hoober, ``Mobile dark patterns,'' \url{https://www.uxmatters.com/mt/archives/2019/11/mobile-dark-patterns.php}, 2019.

\bibitem{repo}
``Artifact availability,'' \url{https://github.com/dangyssss/SCOPEAgent}, 2025.

\bibitem{androiddev}
A.~Developers, ``Request runtime permissions,'' \url{https://developer.android.com/training/permissions/requesting?utm\_source=chatgpt.com\&hl=en}, 2025.

\bibitem{Tales2016}
C.~Bösch, B.~Erb, F.~Kargl, H.~Kopp, and S.~Pfattheicher, ``Tales from the dark side: Privacy dark strategies and privacy dark patterns,'' \emph{Proceedings on Privacy Enhancing Technologies}, vol. 2016, p. 237–254, 07 2016.

\bibitem{di2020ui}
L.~Di~Geronimo, L.~Braz, E.~Fregnan, F.~Palomba, and A.~Bacchelli, ``{UI} dark patterns and where to find them: a study on mobile applications and user perception,'' in \emph{Proceedings of the 2020 CHI Conference on Human Factors in Computing Systems}, 2020, pp. 1--14.

\bibitem{Comparative2022}
\BIBentryALTinterwordspacing
J.~Gunawan, A.~Pradeep, D.~Choffnes, W.~Hartzog, and C.~Wilson, ``A comparative study of dark patterns across web and mobile modalities,'' \emph{Proc. ACM Hum.-Comput. Interact.}, vol.~5, no. CSCW2, Oct. 2021. [Online]. Available: \url{https://doi.org/10.1145/3479521}
\BIBentrySTDinterwordspacing

\bibitem{tc260}
N.~T. C.~. on~Cybersecurity of Standardization Administration~of China, ``Guidelines for the use of system permissions in mobile internet applications,'' \url{https://www.tc260.org.cn/front/postDetail.html?id=20200918163359}, 2020.

\bibitem{delphi}
Wikipedia, ``Delphi method,'' \url{https://en.wikipedia.org/wiki/Delphi\_method}, 2025.

\bibitem{Okoli2004Delphi}
C.~Okoli and S.~Pawlowski, ``The delphi method as a research tool: An example, design considerations and applications,'' \emph{Information \& Management}, vol.~42, pp. 15--29, 12 2004.

\bibitem{Varona_2021}
\BIBentryALTinterwordspacing
D.~Varona and L.~F. Capretz, ``Using the delphi method for model for role assignment in the software industry,'' in \emph{IECON 2021 – 47th Annual Conference of the IEEE Industrial Electronics Society}.\hskip 1em plus 0.5em minus 0.4em\relax IEEE, Oct. 2021, p. 1–7. [Online]. Available: \url{http://dx.doi.org/10.1109/IECON48115.2021.9589957}
\BIBentrySTDinterwordspacing

\bibitem{su132112019}
\BIBentryALTinterwordspacing
E.~Głuszek, ``Use of the e-delphi method to validate the corporate reputation management maturity model (cr3m),'' \emph{Sustainability}, vol.~13, no.~21, 2021. [Online]. Available: \url{https://www.mdpi.com/2071-1050/13/21/12019}
\BIBentrySTDinterwordspacing

\bibitem{frida}
Frida, ``Frida: Dynamic instrumentation toolkit for developers, reverse-engineers, and security researchers.'' \url{https://frida.re/docs/home/}, 2025.

\bibitem{agentsdk}
OpenAI, ``Openai agents sdk,'' \url{https://openai.github.io/openai-agents-python/}, 2025.

\bibitem{yang2022describectx}
S.~Yang, Y.~Wang, Y.~Yao, H.~Wang, Y.~Ye, and X.~Xiao, ``Describectx: context-aware description synthesis for sensitive behaviors in mobile apps,'' in \emph{Proceedings of the 44th International Conference on Software Engineering (ICSE 2022)}, 2022, pp. 685--697.

\bibitem{11023406}
\BIBentryALTinterwordspacing
L.~Wang, D.~Wang, S.~Pan, Z.~Jiang, H.~Wang, and Y.~Wang, ``{ A Big Step Forward? A User-Centric Examination of iOS App Privacy Report and Enhancements },'' in \emph{2025 IEEE Symposium on Security and Privacy (SP)}.\hskip 1em plus 0.5em minus 0.4em\relax Los Alamitos, CA, USA: IEEE Computer Society, 2025, pp. 4210--4228. [Online]. Available: \url{https://doi.ieeecomputersociety.org/10.1109/SP61157.2025.00223}
\BIBentrySTDinterwordspacing

\bibitem{tesseract}
``Tesseract ocr,'' \url{https://github.com/tesseract-ocr/tesseract}, 2025.

\bibitem{droidbot}
DroidBot, ``A lightweight test input generator for android.'' \url{https://github.com/honeynet/droidbot}, 2023.

\bibitem{fastbot}
ByteDance, ``Fastbot-android open source handbook,'' \url{https://github.com/bytedance/Fastbot_Android}, 2023.

\bibitem{privacylabel}
A.~Inc., ``Transparency is the best policy,'' \url{https://www.apple.com/hk/en/privacy/labels/}, 2020.

\bibitem{att}
------, ``If an app asks to track your activity,'' \url{https://support.apple.com/en-hk/102420}, 2021.

\bibitem{APR}
------, ``About app privacy report,'' \url{https://support.apple.com/en-sg/102188}, 2021.

\bibitem{datasafety}
G.~Play, ``Provide information for google play's data safety section,'' \url{https://support.google.com/googleplay/android-developer/answer/10787469?hl=en}, 2022.

\bibitem{dashboard}
------, ``Manage permissions from the privacy dashboard,'' \url{https://support.google.com/android/answer/13530434?visit\_id=63852 6274467900320-3285927032\&p=privacy\_dashboard\&rd=1}, 2022.

\bibitem{zhang2022usable}
S.~Zhang, Y.~Feng, Y.~Yao, L.~F. Cranor, and N.~Sadeh, ``How usable are ios app privacy labels?'' \emph{Proceedings on Privacy Enhancing Technologies}, 2022.

\bibitem{koch2022keeping}
S.~Koch, M.~Wessels, B.~Altpeter, M.~Olvermann, and M.~Johns, ``Keeping privacy labels honest,'' \emph{Proceedings on Privacy Enhancing Technologies}, 2022.

\bibitem{degiulioAskAppNota}
A.~DeGiulio, ````{{Ask App Not}} to {{Track}}'': {{The Effect}} of {{Opt-In Tracking Authorization}} on {{Mobile Privacy}}.''

\bibitem{wu2025transparency}
X.~Wu, L.~Hu, E.~Zeng, H.~Habib, and L.~Bauer, ``Transparency or information overload? evaluating users' comprehension and perceptions of the ios app privacy report.'' in \emph{NDSS}, 2025.

\bibitem{khandelwal2024unpacking}
R.~Khandelwal, A.~Nayak, P.~Chung, and K.~Fawaz, ``Unpacking privacy labels: A measurement and developer perspective on google's data safety section,'' in \emph{33rd USENIX Security Symposium (USENIX Security 24)}, 2024, pp. 2831--2848.

\bibitem{baalous2025detecting}
R.~Baalous, A.~Althobaiti, D.~Alyoubi, R.~Alzahrani, and M.~Aljohani, ``Detecting the inconsistency between android apps’ data collection and google play’s data safety using static analysis,'' \emph{Cybernetics and Information Technologies}, vol.~25, no.~1, 2025.

\bibitem{gray2018dark}
\BIBentryALTinterwordspacing
C.~M. Gray, Y.~Kou, B.~Battles, J.~Hoggatt, and A.~L. Toombs, ``The dark (patterns) side of ux design,'' in \emph{Proceedings of the 2018 CHI Conference on Human Factors in Computing Systems}, ser. CHI '18.\hskip 1em plus 0.5em minus 0.4em\relax New York, NY, USA: Association for Computing Machinery, 2018, p. 1–14. [Online]. Available: \url{https://doi.org/10.1145/3173574.3174108}
\BIBentrySTDinterwordspacing

\end{thebibliography}

\end{document}